\documentclass[twocolumn,graphics,prl,aps,floats,superscriptaddress,showpacs]{revtex4}
\usepackage{graphicx}% Include figure files%\usepackage{dcolumn}% Align table columns on decimal point
\usepackage{amsfonts}
\usepackage{amsmath}
\usepackage{graphicx}

\def\be{\begin{equation}}
\def\ee{\end{equation}}
\def\bea{\begin{eqnarray}}
\def\eea{\end{eqnarray}}

\begin{document}

%%%%%%%%%%%%%%%%%%%%%%%%%%%%%%%%%%%%%%%%%%%%%%%%%%%%%%%%%%%%%%%%%%%%%%%
%%%%%%%%%%%%  BEGIN TITLE %%%%%%%%%%%%%%%%%%%%%%%%%%%%%%%%%%%%%%%%%%%%%
%%%%%%%%%%%%%%%%%%%%%%%%%%%%%%%%%%%%%%%%%%%%%%%%%%%%%%%%%%%%%%%%%%%%%%%

\title{Measuring the Decoherence of a Quantronium Qubit with the Cavity
Bifurcation Amplifier}
\author{M. Metcalfe}
\author{E. Boaknin}
\author{V. Manucharyan}
\author{R. Vijay}
\author{I. Siddiqi}
\author{C. Rigetti}
\author{L. Frunzio}
\author{R. J. Schoelkopf}
\author{M. H. Devoret}
\affiliation{Department of Physics and Applied Physics, Yale University, New Haven, CT,
USA}
\date{\today}

\begin{abstract}
\begin{centering}
Dispersive readouts for superconducting qubits have the advantage of speed and minimal invasiveness.
We have developed such an amplifier, the Cavity
Bifurcation Amplifier (CBA) \cite{etienne}, and applied it to the readout of the quantronium
qubit \cite{vion}. It consists of a Josephson junction embedded
in a microwave on-chip resonator. In contrast with the Josephson bifurcation amplifier
\cite{siddiqi-qm}, which has an on-chip capacitor shunting a junction, the
resonator is based on a simple coplanar waveguide imposing a pre-determined frequency and
whose other RF characteristics like the quality factor are easily controlled and optimized.
Under proper microwave irradiation conditions, the CBA has two metastable states.
Which state is adopted by the CBA depends on the state of a quantronium qubit coupled to the CBA's junction. Due to the MHz repetition rate and large signal to noise ratio we can show directly that the
coherence is limited by 1/f gate charge noise when biased at the \textquotedblleft sweet spot\textquotedblright~-- a
point insensitive to first order gate charge fluctuations. This architecture lends itself
to scalable quantum computing using a multi-resonator chip with multiplexed readouts. 
\end{centering}
\end{abstract}

\maketitle

%%%%%%%%%%%%%%%%%%%%%%%%%%%%%%%%%%%%%%%%%%%%%%%%%%%%%%%%%%%%%%%%%%%%%%%%%%%
%%%%%%%%%%%%%%%% BEGIN INTRO%%%%%%%%%%%%%%%%%%%%%%%%%%%%%%%%%%%%%%%%%%%%%%%
%%%%%%%%%%%%%%%%%%%%%%%%%%%%%%%%%%%%%%%%%%%%%%%%%%%%%%%%%%%%%%%%%%%%%%%%%%%

\section{Introduction}

Quantum circuits based on Josephson junctions are
candidates for the fundamental building block of a quantum computer - a quantum
bit, or qubit \cite{qcircuits}. Several implementations have been tested \cite
{vion,Delft-echo,Semba,Clarke,Rob,Martinis,qlab-JQ}, which may be distinguished by the variable
controlling the state of the qubit (charge, flux or phase) and the technique used for the readout.
These systems can be individually addressed, controlled and read, making them some of the most
advanced solid state qubits.

\begin{figure}[th]
\begin{center}
\includegraphics{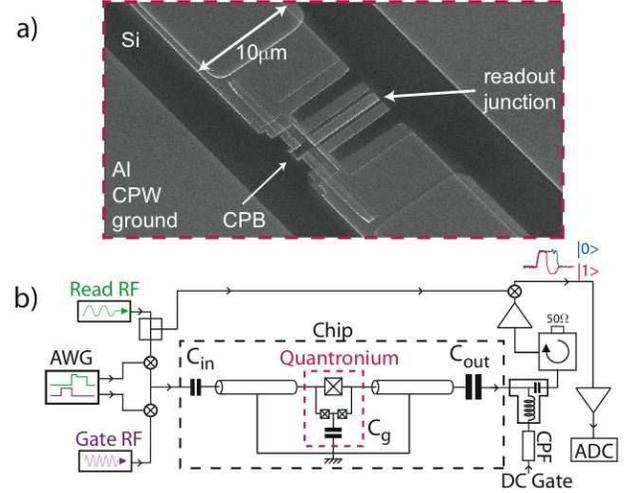}
\end{center}
\caption{\textbf{(a)} Scanning electron micrograph of a duplicate of the
measured device, both of which are fabricated on the same chip simultaneously. \textbf{(b)}
Schematic of our sample and its environment. The sample consists of a split Cooper pair box
whose loop is interrupted with a large junction (quantronium). The state of the quantronium
alters the switching probability $P_{01}$ of the CBA between its two metastable states, which
can be inferred from the amplitude and phase change of transmitted microwave pulses through the device.
}
\label{devicefig}
\end{figure}

An enabling characteristic of many superconducting qubits is the existence of an optimal bias point
where the qubit is immune to first order to fluctuations in external control parameters.
Nonetheless, even systems operated at this \textquotedblleft sweet spot\textquotedblright ~have
 coherence times limited by second order fluctuations of the external control parameters \cite{schon}.
 In this paper we describe measurements of a quantronium qubit operated at this optimal bias point \cite{vion, qlab-JQ} employing the newly developed Cavity Bifurcation Amplifier (CBA) -- a fast, dispersive, scalable readout system based
on a driven non-linear superconducting resonator \cite{etienne,vladimir}. With this architecture
we can measure the fluctuations in the quantronium's coherence on time scales as short as a second,
allowing us to probe the 1/f property of charge noise on these scales. These measurements confirm previous studies
of the limitations of coherence times in charge qubits \cite{schon, Nak-noise} and casts additional light on the
fluctuating character of $T_{2}$ itself and how its value depends on the measurement protocol.

The non-linearity of the CBA is provided by a Josephson junction placed in the center of a $\lambda $/2 on-chip
coplanar waveguide resonator with both an input and output coupling
capacitor playing the role of Fabry-Perot cavity mirrors (see Fig \ref{devicefig}.)
When driven with a microwave signal at frequency $\nu _{d}$ such that $\nu _{d}<\nu _{0}\left(
1-\sqrt{3}/Q\right) $, where $\nu _{0}$ is the resonator small oscillation
natural frequency and $\ Q$ the loaded quality factor, this system can have two
dynamical metastable states which differ by their oscillation amplitude and
phase. As the driving power $P$ is ramped past the bifurcation power $P_{b}$
, the CBA switches from the state of low amplitude to the state with high
amplitude. We detect the state of the non-linear oscillator by monitoring
the amplitude and phase change of the microwave signal transmitted by the
resonator. In parallel with the CBA's junction we place a split Cooper pair box (SCPB), a
circuit known as the quantronium, giving
the resonator two bifurcation powers $P_{b}^{\left\vert 0\right\rangle }$ and $%
P_{b}^{\left\vert 1\right\rangle }<P_{b}^{\left\vert 0\right\rangle }$
depending on the state of the qubit, $\left\vert 0\right\rangle $ or $\left\vert 1\right\rangle $.
The two qubit states are mapped into the two metastable states of the
CBA by ramping quickly the power $P$ to a level intermediate between
$P_{b}^{\left\vert 0\right\rangle }$ and $P_{b}^{\left\vert1\right\rangle }$.
If the quantronium qubit is in $\left\vert 1\right\rangle
$, the CBA will switch to the high oscillating state, whereas if it is in $\left\vert 0\right\rangle $
the CBA will remain in the low oscillating state.

This readout has the advantage of being non-dissipative as the readout junction never switches
into the normal state, unlike the original DC-biased quantronium readout
\cite{vion}. This dispersive readout minimally disturbs the qubit state and since after switching we do not need to wait for quasiparticles to relax, the repetition rate is only limited by the
relaxation time, $T_{1}$, of our qubit and the $Q$ of our resonator. Like the DC readout, the CBA readout
can latch \cite{siddiqi-qm}, allowing enough time for the measurement of the complex amplitude
of the transmitted wave, and therefore excellent signal to noise ratio.
These characteristics were also present in the Josephson bifurcation
amplifier \cite{siddiqi,siddiqi-qm,qlab-JQ}, which implemented a bifurcating
non-linear oscillator using a lumped element capacitor in parallel with the
junction. However this capacitor was fabricated using a Cu/Si$_{3}$N$_{4}$/Al
multilayer structure which was difficult to fabricate and integrate with more
then one qubit. Also the parallel plate
geometry suffered from inherent stray inductive elements.
In contrast, the CBA is fabricated using a simple coplanar waveguide geometry with no
stray elements. The resonance frequency $\nu _{0}$ and the quality factor $Q$
are controlled by the resonator length and output capacitor respectively.
The CBA geometry thus offers the possibility of designing a
multi-resonator chip with multiplexed readouts, which would accommodate up to
10 qubits at once, an important step towards scalable quantum computing. The
present work, in addition to the assessment of 1/f noise in a new architecture,
is a first step in this direction.

%%%%%%%%%%%%%%%%%%%%%%%%%%%%%%%%%%%%%%%%%%%%%%%%%%%%%%%%%%%%%%%%%%%%%%%%%%%
%%%%%%%%%%%%%%%%%% END INTRO%%%%%%%%%%%%%%%%%%%%%%%%%%%%%%%%%%%%%%%%%%%%%%%
%%%%%%%%%%%%%%%%%%%%%%%%%%%%%%%%%%%%%%%%%%%%%%%%%%%%%%%%%%%%%%%%%%%%%%%%%%%

%%%%%%%%%%%%%%%%%%%%%%%%%%%%%%%%%%%%%%%%%%%%%%%%%%%%%%%%%%%%%%%%%%%%%%%%%%%
%%%%%%%%%%%%%% BEGIN DEVICE %%%%%%%%%%%%%%%%%%%%%%%%%%%%%%%%%%%%%%%%%%%%%%%
%%%%%%%%%%%%%%%%%%%%%%%%%%%%%%%%%%%%%%%%%%%%%%%%%%%%%%%%%%%%%%%%%%%%%%%%%%%

\section{Sample fabrication and characterization}

The resonator is initially fabricated using photolithography on a bare Si
wafer \cite{Luigi}. A LOR5A/S1813 optical resist bilayer is used and the
development is optimized to have at least 50nm of undercut beneath the S1813 to
avoid wavy edges and to obtain a sloped edge on the resonator. This sloped edge
 is obtained by evaporating a 200nm thick Al layer onto the sample at 0.2 nm/s with
 an angle of $5^{\circ}$ and with a stage rotation of $10^{\circ}$/s.
  Next the quantronium is fabricated using electron beam
lithography. We use a MMA/PMMA resist bilayer and the Dolan bridge double
angle evaporation technique to fabricate our junctions \cite{Dolan}. For
this sample the split Cooper pair box is fabricated first inside the
resonator, followed by the readout junction in a separate step using new e-beam
resist. Using a hollow cathode Ar ion gun we obtain an ohmic contact between the two e-beam
layers and the resonator.

For the qubit sample described in this paper, we have been working with a
linear regime resonance frequency $\nu _{0}=9.64~\mathrm{GHz}$ and a $Q$ of
160. An on-chip twin (Fig \ref{devicefig}a) of the SCPB of the quantronium had a normal state
resistance of $15~\mathrm k \Omega $, with small junction areas of $0.05~\mathrm \mu $m$^{2}$.
The usual gate line of the quantronium is absent in our design since we
use the readout lines to access both the RF and DC gate lines, another
advantage of this configuration.

%%%%%%%%%%%%%%%%%%%%%%%%%%%%%%%%%%%%%%%%%%%%%%%%%%%%%%%%%%%%%%%%%%%%%%%%%%%
%%%%%%%%%%%%%%%% END DEVICE %%%%%%%%%%%%%%%%%%%%%%%%%%%%%%%%%%%%%%%%%%%%%%%
%%%%%%%%%%%%%%%%%%%%%%%%%%%%%%%%%%%%%%%%%%%%%%%%%%%%%%%%%%%%%%%%%%%%%%%%%%%

%%%%%%%%%%%%%%%%%%%%%%%%%%%%%%%%%%%%%%%%%%%%%%%%%%%%%%%%%%%%%%%%%%%%%%%%%%%
%%%%%%%%%%%%% BEGIN GATE MODULATIONS %%%%%%%%%%%%%%%%%%%%%%%%%%%%%%%%%%%%%%
%%%%%%%%%%%%%%%%%%%%%%%%%%%%%%%%%%%%%%%%%%%%%%%%%%%%%%%%%%%%%%%%%%%%%%%%%%%

\section{Qubit characterization}

\begin{figure}[th]
\begin{center}
\includegraphics{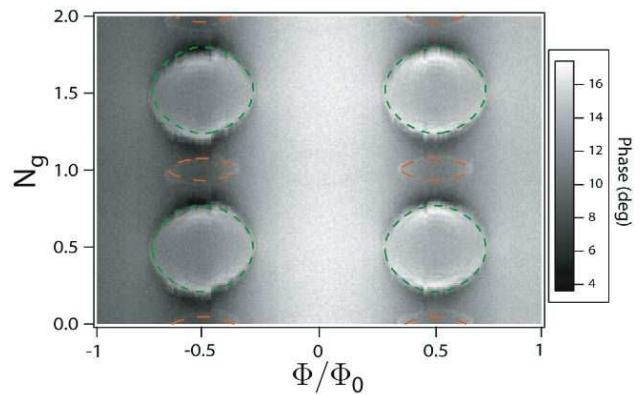}
\end{center}
\caption{Gate charge and flux modulations of our device. We operated in the weakly non-linear mode ($P\ll
P_{b}^{\left\vert 1\right\rangle }$) and monitored the phase (gray-scale) as we varied the applied gate
charge $N_{g}=C_{g}V_{g}/2e$ and flux $\Phi$ ($\Phi_{0}=\hbar/2e$). The large ellipsoidal contours can be interpreted as induced transitions between the energy levels of the qubit at multiples of the
readout frequency. The green fitted lines are transitions between the 0 and
1 energy levels at the readout frequency of $9.64~\mathrm{GHz}$, while the orange fits
are for transitions between the 0 and 2 energy levels at twice the readout
frequency, $19.28~\mathrm{GHz}$}
\label{gatemod}
\end{figure}

We performed gate charge and flux modulations while keeping the
qubit in its ground state, to check that we have flux periodicity and 2e charge
periodicity, as shown in Fig \ref{gatemod}.
The readout was first operated in the weakly non-linear mode ($P\ll
P_{b}^{\left\vert 1\right\rangle }$) and we measured changes in the phase of
the transmitted signal as the gate charge and flux were varied, keeping the
frequency fixed at the maximal phase response point (see Fig \ref{gatemod}).
Apart from a slow background modulation due to the changing susceptibility
of the ground state, we observe sharp contrast on contours of ellipsoidal
shape. These can be interpreted as contours of constant qubit transition
frequency coinciding with the readout frequency or its double, an effect
similar to that observed by Wallraff et al. \cite{Rob}.
Using the well known expressions for the energy levels of the quantronium
\cite{cottet}, we can reproduce the shape of these contours, within the
uncertainty due to the low frequency gate charge and flux noises, and
extract a Jospehson energy of the SCPB $E_{J}$ of $15~\mathrm{GHz}$ and
charging energy $E_{CP}=(2e)^{2}/2C_{\sum}$ of $17~\mathrm{GHz}$, where $C_{\sum}$
is the sum of the capacitances of the junctions in the SCPB and the
gate capacitance.
%At higher drive powers, close to, or in the bifurcation regime, we have observed more
%complex features, involving higher order transitions, and confirming the
%qubit parameters (data not shown).

\begin{figure}[th]
\begin{center}
\includegraphics{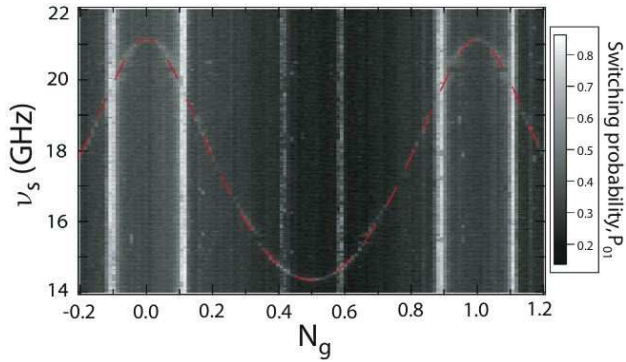}
\end{center}
\caption{Spectroscopy peak as a function of gate charge, $N_{g}$. Staying at zero flux, we measure $P_{01}$ while sweeping $\nu_{s}$ and stepping $N_{g}$. The theoretical fit of the resulting sinusoidal like dependance of the peak with $N_{g}$ is given by the red dashed line with the fit parameters $E_{J}=15.02~\mathrm{GHz}$ and $E_{CP}=17.00~\mathrm{GHz}$. The vertical lines with no $N_{g}$ dependance are the excitations between qubit
energy levels enduced at multiples of the readout frequency, similar to those seen in Fig \protect\ref{gatemod}}
\label{spec}
\end{figure}
To get a more precise measurement of $E_{J}$ and $E_{CP}$, we performed
spectroscopy on the qubit by applying a weakly exciting $1~\mathrm \mu s$ long
pulse, followed by a latching readout pulse.
The switching probability $P_{01}$ between the two metastable states of the CBA
is measured as the spectroscopic frequency $\nu_{s}$ is swept for each gate charge step, at zero flux.
Due to the crosstalk between the readout and the qubit, we have to ensure that we have
zero leakage of spectroscopic power outside our pulse. This is achieved by gating the LO on
the mixers shaping our pulses with a pulse shape a few ns longer then the spectroscopic pulse.
As a function of frequency, we find a peak in switching probability, whose position
varies with gate charge with the expected sinusoidal-like shape shown in Fig \ref{spec}.
The theoretical fit, shown in red, refines the previous determination of $E_{J}$ and $E_{CP} $ to
the values $E_{J}=15.02~\mathrm{GHz}$ and $E_{CP}=17.00~\mathrm{GHz}$. Zooming in to the double
\textquotedblleft sweet spot\textquotedblright , $N_{g}=0.5$, $\Phi /\Phi _{0}=0$, where the
qubit is immune to charge and flux noise to first order, we measure a
Lorentzian spectroscopic peak of width $\Delta \nu _{01}=2.6~\mathrm{MHz}$ and a
Larmor frequency $\nu _{01}=14.35~\mathrm{GHz}$. This gives a dephasing time, $
T_{2}=1/\pi \Delta \nu _{01}$ of $120~\mathrm{ns}$. However, large charge jumps move
the biasing point off the \textquotedblleft sweet spot\textquotedblright~causing the linewidth to be widened.
More accurate measurements of $T_{2}$ will be obtained from Ramsey fringes
where we can follow the variation of $T_{2}$ with time.
%%%%%%%%%%%%%%%%%%%%%%%%%%%%%%%%%%%%%%%%%%%%%%%%%%%%%%%%%%%%%%%%%%%%%%%%%%%%
%%%%%%%%%%%%%%%% END GATE MODULATIONS %%%%%%%%%%%%%%%%%%%%%%%%%%%%%%%%%%%%%%
%%%%%%%%%%%%%%%%%%%%%%%%%%%%%%%%%%%%%%%%%%%%%%%%%%%%%%%%%%%%%%%%%%%%%%%%%%%%

%%%%%%%%%%%%%%%%%%%%%%%%%%%%%%%%%%%%%%%%%%%%%%%%%%%%%%%%%%%%%%%%%%%%%%%%%%%%
%%%%%%%%%%%%%%% BEGIN QUBIT PERFORMANCE %%%%%%%%%%%%%%%%%%%%%%%%%%%%%%%%%%%%
%%%%%%%%%%%%%%%%%%%%%%%%%%%%%%%%%%%%%%%%%%%%%%%%%%%%%%%%%%%%%%%%%%%%%%%%%%%%

\section{Qubit manipulation and coherence experiments}

Once the qubit parameters are known, we can perform
experiments on the qubit to determine the qubit's
quality in terms of its energy relaxation time $T_{1}$ and dephasing time
$T_{2}$. An essential part of these experiments is the need to control
the state of the qubit precisely. This is achieved by applying a
microwave pulse to the gate with rectangular envelope, of amplitude $A$, and time
length $\tau $. In a frame rotating at the gate pulse carrier frequency, $
\nu _{s}$, the qubit state then precesses at a frequency, $\nu _{p}$ given
by \cite{cottet}:
\begin{equation}
\nu _{p}=\left[ \left( \nu _{Rabi}\right) ^{2}+(\nu _{01}-\nu _{s})^{2}
\right] ^{1/2}  \label{prec}
\end{equation}

\begin{figure}[th]
\begin{center}
\includegraphics{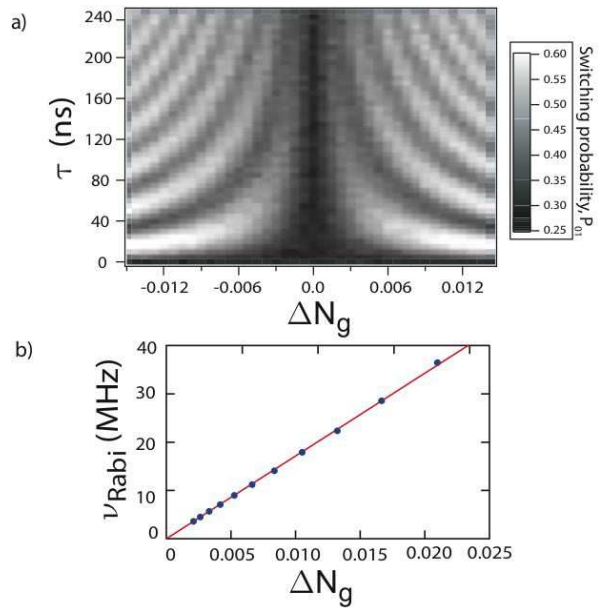}
\end{center}
\caption{\textbf{(a)} Rabi oscillations in the switching probability $P_{01}$, as a function of gate
charge modulation $\Delta N_{g}$ and gate pulse time
length $\protect\tau $. $\Delta N_{g}$ is calculated from the Rabi pulse envelope
voltage A reaching the sample through the attenuation in the input lines
and is plotted in terms of Cooper pairs, $\Delta N_{g}=C_{g}A/2e$.
Oscillations in the switching probability $P_{01}$ are seen with both $\Delta N_{g}$
and $\protect\tau $. \textbf{(b)} Fitted Rabi frequency $\nu_{Rabi}$ vs $\Delta N_{g}$.
As expected from a two-level system, $\nu_{Rabi}$ scales linearly with $\Delta N_{g}$.}
\label{rabi}
\end{figure}

\begin{figure*}[th]
\begin{center}
\includegraphics{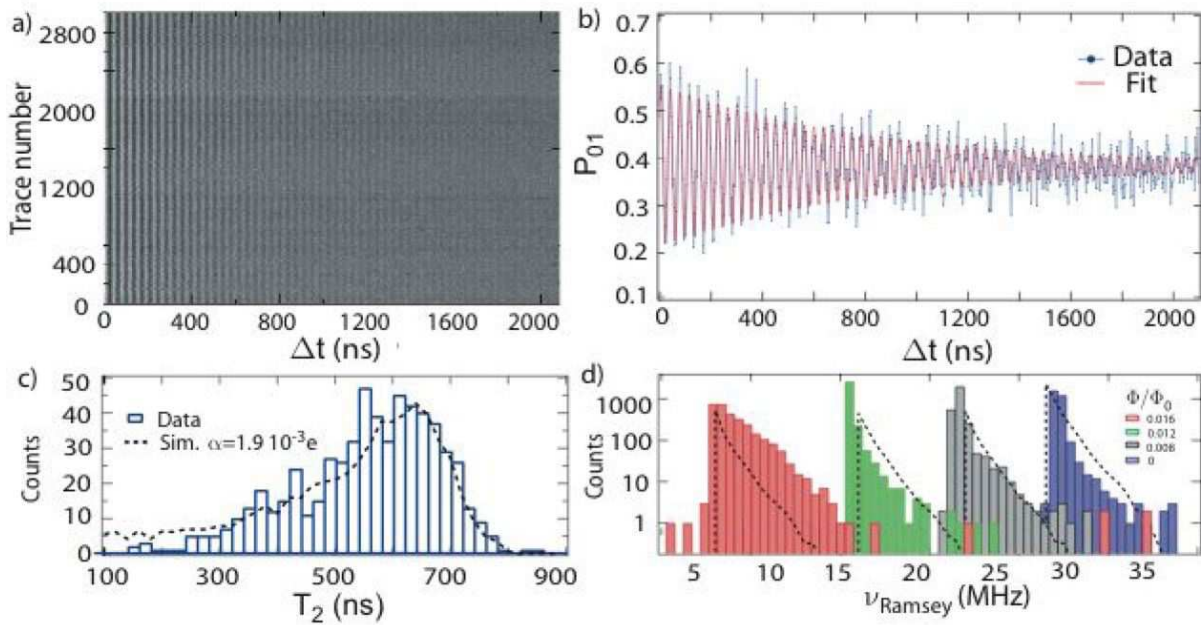}
\end{center}
\caption{\textbf{(a)} 3000 Ramsey oscillations as a function of free evolution time $\Delta t$.
Each trace is $2.1~\mathrm{\mu s}$ long with $3~\mathrm{ns}$ per step. They each take $0.35~\mathrm{s}$ to acquire. We can see visually the variation of $T_{2}$ for the different Ramsey fringes by noticing the variation in contrast in the fringes near $2~\mathrm{\mu s}$.
\textbf{(b)} Sample data fit. We average 5 of the acquired Ramsey traces shown in (a) and fit
to a decaying sinusoid to extract the $T_{2}$ which is then plotted in (c). For
this case we have a coherence time of $842~\mathrm{ns}$ and a Ramsey frequency of
$26.9~\mathrm{MHz}$. \textbf{(c)} Distribution of $T_{2}$ for 3000 of the Ramsey
traces (600 fits). The black dashed line is the result of
a simulation of the free evolution decay of the Ramsey
fringes with 1/f noise fluctuations on the gate, $S_{q}(\protect\omega )=
\protect\alpha ^{2}/|\protect\omega |$. In the simulation we used 10 times more points
compared to the data to obtain a smoother curve. \textbf{(d)} Corresponding
distribution of Ramsey frequencies at four different flux biasing points. Each
distribution has 3000 Ramsey traces. The blue histogram corresponds to the data in
(a), (b) and (c). The Ramsey frequency is extracted from
the position of the maximum of the power spectral density of each decaying
sinusoid. The distributions are lopsided to higher frequencies as would be expected from
fluctuations in the gate charge around our operating point at the \textquotedblleft sweet spot\textquotedblright.
The dashed line is the expected distribution assuming the same 1/f charge noise as in (c).
%As we move to larger flux the distributions tend to broaden which could be an indication
%of an increase in flux noise as we move away from the \textquotedblleft sweet spot\textquotedblright~in flux.
}
\label{ramsey}
\end{figure*}

where the Rabi frequency \cite{rabi} $\nu _{Rabi}=2E_{CP}\Delta
N_{g}\left\langle 0\left\vert N\right\vert 1\right\rangle /h$ involves the
gate modulation ampitude in units of Cooper pairs $\Delta N_{g}=C_{g}A/2e$
and charge operator matrix element between the excited and ground states $
\left\langle 0\left\vert N\right\vert 1\right\rangle $. For $A=0$ we have
free evolution at the Ramsey frequency \cite{ramsey} $\nu _{Ramsey}=|\nu
_{01}-\nu _{s}|$.

We begin by measuring the Rabi oscillations. The pulse sequence protocol
involves a gate pulse with fixed $\nu _{s}=$ $\nu _{01}$ and varying
amplitude, $A$, and time length $\tau $, followed by a latching readout pulse.
The sequence is repeated $10^{4}$ times to measure the switching probability. The
oscillations of the switching probability as a function of $\tau $ and $A$
are plotted in Fig \ref{rabi}. The extracted frequency $\nu _{Rabi}$ scales
linearly $\Delta N_{g}$ (Fig \ref{rabi}b) as expected from a two-level
system. From the position of the first maximum of the Rabi oscillations we can calibrate the
pulse time length needed for a $\pi $-pulse, which drives the qubit from the
ground state to the excited state. Using this $\pi $-pulse we measure the
exponential decay of the population of the excited state (data not shown)
and obtain the relaxation time $T_{1}$ which, during the course of an
experimental run, varied between $1.4~\mathrm{\mu s}$ and $1.8~\mathrm{\mu s}$, with a
persistence time of a few seconds. These values are comparable to the
results of Vion et al. \cite{vion} and Siddiqi et al. \cite{qlab-JQ}.

The maximum constrast measured is about 60\%, lower then the maximum contrast
of over 99.9\%, calculated for the ideal case of a non-relaxing qubit given the
measured parameters of the resonator. This loss can be attributed to three main sources.
First the qubit relaxes before the readout takes place,
because of its finite $T_{1}$, resulting in a 10\% loss in contrast.
Second the qubit relaxes to the ground state as the readout voltage
approaches the bifurcation voltage, resulting in a further 25\% loss in contrast.
Other measurements (not described here) have suggested that this loss in contrast
could be from Stark shifting the qubit to lower frequencies during readout, where
it can come in resonance with spurious transitions, possibly
due to defects in the substrate or in the tunnel barrier.
The remaining loss could be accounted for by the fact that the transition between the two
oscillating states of the CBA is broadened by more than a factor of 5 from that expected.

To measure the coherence time, $T_{2}$, we follow a different pulse protocol
in which we apply two $\pi /2$ pulses separated by a free evolution period
of length $\Delta t$, followed by a readout measurement. An example of the
resulting Ramsey fringes is shown in Fig \ref{ramsey}b. By fitting to an
exponentially decaying sinusoid we extract $T_{2}$.
The frequency of Ramsey fringes is well fitted by the absolute value of the detuning $|\nu _{s}-\nu
_{01}|$, yielding a precise measurement of $\nu _{01}=14.346~\mathrm{GHz}$.
We follow the time evolution of $T_{2}$ and $\nu _{01}$ by recording a Ramsey
fringe every $0.35~\mathrm{s}$ (Fig \ref{ramsey}a). We observe a variation of $T_{2}$ from $150~\mathrm{ns}$
to $850~\mathrm{ns}$. The distribution of $T_{2}$ is an asymmetric bell shaped curve
peaking around $600~\mathrm{ns}$, with a long tail extending down to $150~\mathrm{ns}$ (Fig \ref
{ramsey}c). If we would average over a $17.5~\mathrm{mn}$ period (3000 of the above Ramsey traces)
we would measure an average $T_{2}$ which converges to $500~\mathrm{ns}$, similar to the first Saclay result
\cite{vion} obtained with a qubit with a similar $E_{J}/E_{CP}$. The $T_{2}$
fluctuations are correlated with fluctuations in the Ramsey frequency, which
only fluctuates towards higher frequencies, giving lopsided distributions,
as shown on Fig \ref{ramsey}d. At the \textquotedblleft sweet spot\textquotedblright~where we are working,
variations in gate charge necessarily increase the transition frequency
whereas variations in flux decrease it. Variations in critical current would
supposedly keep the distribution of frequencies more symmetric. We can therefore
conclude that charge noise, not flux noise, is the dominant source of
decoherence in our sample. Furthermore, if we suppose that the charge noise
is Gaussian with a spectral density that has the usually invoked 1/f form \cite{Oneoverf}
given by $S_{q}(\omega )=\alpha ^{2}/|\omega |$, we can check
if our data can be explained by this model. This was carried out by directly
numerically simulating the corresponding variations in transition frequency
and calculating the Ramsey signal in the conditions of the experiment. The
distributions of both the extracted $T_{2}$ and $\nu _{Ramsey} $ are shown by
the dashed lines in Fig \ref{ramsey}c and d. We obtain good agreement with
the data for a noise amplitude of $\alpha =1.9.10^{-3}e$, agreeing with the
range of previously measured values of this noise intensity parameter. To reduce
sensitivity to this charge noise we can make the energy levels of the qubit
almost insensitive to charge by increasing the areas of the junctions in the
SCPB and hence increasing $E_{J}/E_{CP}$. An $E_{J}/E_{CP}$ of 8 could give 
a $T_{2}$  in the ms range and hence this device would be $T_{1}$ limited.

%%%%%%%%%%%%%%%%%%%%%%%%%%%%%%%%%%%%%%%%%%%%%%%%%%%%%%%%%%%%%%%%%%%%%%%%%%%%
%%%%%%%%%%%%%%%%% END QUBIT PERFORMANCE %%%%%%%%%%%%%%%%%%%%%%%%%%%%%%%%%%%%
%%%%%%%%%%%%%%%%%%%%%%%%%%%%%%%%%%%%%%%%%%%%%%%%%%%%%%%%%%%%%%%%%%%%%%%%%%%%

%%%%%%%%%%%%%%%%%%%%%%%%%%%%%%%%%%%%%%%%%%%%%%%%%%%%%%%%%%%%%%%%%%%%%%%%%%%%
%%%%%%%%%%%%%%%%%%%%%% BEGIN CONCLUSION %%%%%%%%%%%%%%%%%%%%%%%%%%%%%%%%%%%%
%%%%%%%%%%%%%%%%%%%%%%%%%%%%%%%%%%%%%%%%%%%%%%%%%%%%%%%%%%%%%%%%%%%%%%%%%%%%

\section{Conclusion}

We have successfully implemented an improved version of the bifurcation amplifier
based on an on-chip CPW resonator as a readout for the quantronium qubit. It
offers ease of fabrication and a larger range of operating parameters ($%
\omega _{0}$, Q) compared to the original JBA implementation \cite{qlab-JQ}.
Using this readout which captures in real time the fluctuations in qubit
parameters, we have demonstrated that the main source of decoherence in our
sample is charge noise. By using a larger $E_{J}/E_{CP}$, we could reduce the
curvature with gate charge of the levels of the Cooper pair box and we
should be able to reduce the charge noise induced decoherence \cite{transmon}.
Furthermore, the CBA geometry is particularly well adapted to the
multiplexing of the simultaneous readout of several qubits. We have started
in this direction by successfully measuring the bifurcation of 5 CBAs with
only one input and one output line. This configuration offers a path for
scaling up of superconducting circuits up to several tens of qubits.

The authors would like to thank D. Esteve, S. Fissette,  J.M. Gambetta,
S. Girvin, D. Prober and D. Vion for useful discussions and assistance.
This work was supported by NSA through ARO grant No. W911NF-05-01-0365, the
Keck foundation, and the NSF through grant No. DMR-032-5580. L. Frunzio acknowledges
partial support from CNR-Istituto di Cibernetica, Pozzuoli, Italy.

%%%%%%%%%%%%%%%%%%%%%%%%%%%%%%%%%%%%%%%%%%%%%%%%%%%%%%%%%%%%%%%%%%%%%%%%%%%%
%%%%%%%%%%%%%%%%%%%%%%%% END CONCLUSION %%%%%%%%%%%%%%%%%%%%%%%%%%%%%%%%%%%%
%%%%%%%%%%%%%%%%%%%%%%%%%%%%%%%%%%%%%%%%%%%%%%%%%%%%%%%%%%%%%%%%%%%%%%%%%%%%

%%%%%%%%%%%%%%%%%%%%%%%%%%%%%%%%%%%%%%%%%%%%%%%%%%%%%%%%%%%%%%%%%%%%%

%%%%%%%%%%%%%%%%%%%%%%%%%%%%%%%%%%%%%%%%%%%%%%%%%%%%%%%%%%%%%%%%%%%%%%%%%%%%
%%%%%%%%%%%%%%%%%%%%%%% BEGIN BIBLIOGRAPHY %%%%%%%%%%%%%%%%%%%%%%%%%%%%%%%%%
%%%%%%%%%%%%%%%%%%%%%%%%%%%%%%%%%%%%%%%%%%%%%%%%%%%%%%%%%%%%%%%%%%%%%%%%%%%%

%%%%%%%%%%%%%%%%%%%%%%%%%%%%%%%%%%%%%%%%%%%%%%%%%%%%%%%%%%%%%%%%%%%%%

\end{document}